\documentclass{aa}
\usepackage{graphics}
\usepackage{amssymb}

\begin{document}

\thesaurus{06 (06.18.1; 02.03.1; 03.13.2; 03.13.6 )}

\title{Determination of fractal dimensions of solar radio bursts}

\author{A. Veronig \inst{1}
 \and M. Messerotti \inst{2}
 \and A. Hanslmeier \inst{1} }

\offprints{A. Veronig}

\institute{Institute of Astronomy, University of Graz,
 Universit\"atsplatz 5, A--8010 Graz, Austria
 \and Trieste Astronomical Observatory, Via G.B. Tiepolo 11,
 I--34131 Trieste, Italy }

\date{Received 19 July 1999 / Accepted 7 March 2000}

\maketitle

\begin{abstract}
We present a dimension analysis of a set of solar type~I storms
and type~IV events with different kind of fine structures, recorded
at the Trieste Astronomical Observatory. The signature of such types
of solar radio events is highly structured in time. However,
periodicities are rather seldom, and linear mode theory can provide
only limited interpretation of the data. Therefore, we performed an
analysis based on methods of the nonlinear dynamics theory.

Additionally to the commonly used correlation dimension, we also
calculated local pointwise dimensions. This alternative approach is
motivated by the fact that astrophysical time series represent
real-world systems, which cannot be kept in a controlled state and
which are highly interconnected with their surroundings. In such systems
pure determinism is rather unlikely to be realized, and therefore a
characterization by invariants of the dynamics might probably be inadequate.

In fact, the outcome of the dimension analysis does not give hints for
low-dimensional determinism in the data, but we show that, contrary to the
correlation dimension method, local dimension estimations can give physical
insight into the events even in cases in which pure determinism cannot be
established. In particular, in most of the analyzed radio events nonlinearity
in the data is detected, and the local dimension analysis provides a basis for
a quantitative description of the time series, which can be used to
characterize the complexity of the related physical system in a comparative and
non-invariant manner.

In this frame, the degree of complexity we inferred for type~I storms is on
the average lower than that relevant to type~IV events. For the type~IV events
significant differences occur with regard to the various subtypes, whereas
pulsations and sudden reductions can be described by distinctly lower values
than spikes and fast pulsations.

\end{abstract}

\section{Introduction}

Nonlinear time series analysis based on the theory of
deterministic chaos has turned out to be a powerful tool in
understanding complex dynamics from measurements and observational
time series. In particular it can provide
descriptions and interpretations for irregular times series, which
nevertheless might not be governed by a stochastic physical
process and which are only poorly understood by linear methods.
A number of recent reviews and conference proceedings shows the
great interest in the field of nonlinear time series analysis
(see, for instance, Grassberger et al. \cite{GrassbergerEtal91};
Casdagli \& Eubank \cite{CasdagliEubank92}; Weigend \& Gershenfeld
\cite{WeigendGershenfeld93}; Kugiumtzis \cite{KugiumtzisEtal94a},
\cite{KugiumtzisEtal94b}; Abarbanel \cite{Abarbanel96}; Kantz \&
Schreiber \cite{KantzSchreiber97}; Schreiber \cite{Schreiber99}).

Since the development of chaos theory, it is well known that even
simple dynamical systems, described by few nonlinear differential
equations, can reveal a complex and quasi-irregular behavior.
A central concept to characterize such systems is the
so-called {\em attractor}. Under the dynamics of a deterministic
system the trajectories do not cover the whole phase space, but,
after all transient phenomena have faded out, converge to a subset
of the phase space, the attractor. The attractor itself is invariant
to the dynamical evolution. Simple examples of attractors are
fixed points and limit cycles. However, when the related dynamical
system is {\em chaotic}, the attractor can have a complex
geometry with a {\em fractal}, i.e., non-integer, dimension.
Different invariant parameters exist to describe the geometry
and the dynamics of an attractor, such as dimensions, Lyapunov
exponents and entropies. Besides the fractal geometry, chaotic
systems have the striking property that initially neighboring
trajectories diverge exponentially under the dynamics, and the
growth rate is given by the {\em Lyapunov exponent}. This
phenomenon results from the folding and stretching of the
trajectories under the dynamics, the folding leading to the
convergence of the trajectories to the attractor and the
stretching to the divergence in certain directions. While the
average stretching rate is given by the Lyapunov exponent, the
loss of information due to the folding is quantified by the {\em
entropy}. Lyapunov exponents and entropies characterize the
dynamics on the attractor, and the {\em dimension}
characterizes its geometry. The physical meaning of the dimension
of an attractor is that it corresponds to the degree of freedom of
the related dynamical system, i.e., in a deterministic case, the
minimum number of ordinary differential equations needed to fully
describe the system. Deterministic systems are characterized by a
finite dimension. Deterministic chaotic systems have the additional
characteristic that their dimension is fractal.
Contrary to that, a stochastic system is characterized by an
infinite dimension, indicating its infinite degree of freedom.
Therefore, the determination of the dimension of an attractor
enables to discriminate whether a dynamical system is
deterministic or stochastic.

Previous papers exist concerning the investigation of fractal dimensions
of solar radio bursts. It has to be noted, that the time series used
by the different authors are not directly comparable as they represent
different types of radio events. Kurths \& Herzel (\cite{KurthsHerzel86},
\cite{KurthsHerzel87}),  Kurths \& Karlick$\acute{\rm y}$ (\cite{KurthsKarlicky89}),
and Kurths et al. (\cite{KurthsEtal91}) analyzed decimetric pulsations
and ascertained finite dimension values. Contrary to that,
Isliker (\cite{Isliker92b}) and Isliker \& Benz (\cite{IslikerBenz94a},
\cite{IslikerBenz94b}) investigated different types of solar radio bursts
in the metric~(m) and decimetric~(dm) wavelength range (type~I storms,
type~II bursts, type~III bursts, type~IV events, and narrowband
spikes), which did not reveal any hints for low-dimensional
determinism.\footnote{The reported finite dimension for one of
the analyzed narrowband spike events in Isliker (\cite{Isliker92b}) was
revised in a later paper (Isliker \& Benz \cite{IslikerBenz94a}).}

However, these investigations rely all on the correlation dimension method. The
present paper additionally introduces a complementary dimension analysis, motivated
by the fact that solar radio bursts represent real-world systems, which implies
some major restrictions. First, the time series cannot be expected to be stationary,
and second, pure determinism is rather unlikely to be realized. Therefore we do not
only concentrate on the usual way of looking at the problem: ``Does the analyzed
time series represent a deterministic or a stochastic system?" but in particular
focus the question: ``What statistical description can be extracted from a dimension
analysis of the time series?" With such refined formulation of the problem, we try
to make use of the concepts and tools of nonlinear time series analysis even in
cases in which the determination of invariants of the dynamics, as, e.g., attractor
dimensions, possibly fails.

The paper is structured as follows. Sect.~\ref{Methods} explains the used
methods and discusses critical points in the determination of fractal dimensions
from time series. In Sect.~\ref{DataSets} the investigated data sets are
characterized and the analysis procedure is described. Sect.~\ref{Results}
presents the results of the dimension analysis, which are discussed in
Sect.~\ref{Discussion}. Finally, the conclusions are drawn in Sect.~\ref{Conclusion}.

\section{Methods  \label{Methods}}

\subsection{Phase space reconstruction}

Generally, not all relevant parameters of the dynamics of a system are
measured during an observation but only a one-dimensional time series
is given. To reconstruct the phase space of the related dynamical system,
techniques have to be applied to unfold the multi-dimensional attractor
from a scalar time series. By the technique of time delayed coordinates
(Takens \cite{Takens81}), from a given one-dimensional time
series~$\{x(t_i)\}$ an $m$-dimensional phase space~$\{\mbox{\boldmath $\xi$}_i\}$
is built up by the prescription
\begin{equation}
 \mbox{\boldmath $\xi$}_i = \{ x(t_{i}), \; x( t_{i}+\tau), \; \ldots \; ,x(t_{i}+(m-1)\tau) \}\; ,
\end{equation}
where $\tau$ is the time delay.

According to the embedding theorem of Takens (\cite{Takens81}), the embedding
of the attractor in the $m$-dimensional reconstructed phase space can be
ensured if \mbox{$m \ge (2D+1)$}, with~$D$ the dimension of the original phase
space. For time series of infinite length and accuracy, $\tau$ can be chosen
in a more or less arbitrary way without affecting the results. However, in
practice not every value for the time delay~$\tau$ will be suitable. Too
small~$\tau$ will build up coordinates which are too strongly correlated, while
for large~$\tau$ the vector components show no causal connection. The choice
of the time delay~$\tau$ in the reconstruction of the phase space usually
strongly affects the quality of the analysis, and different procedures have
been worked out to ensure a proper choice of~$\tau$. The most prominent methods
use the auto-correlation time ($1/e$~decay time, first zero crossing,
first minimum) or the first minimum of the mutual information (Fraser \&
Swinney \cite{FraserSwinney86}).

One advantage of the {\em mutual information} over the auto-correlation
function is that it takes into account nonlinear properties of the data.
The mutual information is based on the Shannon entropy (Shannon \& Weaver
\cite{ShannonWeaver62}), and gives the information about the state of a
system at time $t+\tau$ that we already possess if we know the state at
time~$t$. The choice of the first minimum of the mutual information for
the time delay~$\tau$ is motivated by the fact that two successive delay
coordinates should be as independent as possible without making~$\tau$
too large.

\subsection{Correlation dimension}

The correlation dimension is one out of many definitions of fractal dimensions,
and was introduced by Grassberger \& Procaccia (\cite{GrassbergerProcaccia83a},
\cite{GrassbergerProcaccia83b}) to determine fractal dimensions from time series.
The correlation dimension is based on distance measurements of points in phase
space. Therefore, as first step, from the time series~$\{x(t_i)\}$ the phase
space vectors~$\{\mbox{\boldmath $\xi$}_i\}$ have to be constructed. With the
reconstructed vectors, the correlation integral~$C(r)$ can be calculated, which
is given by the normalized number of pairs of points within a distance~$r$. As
the correlation dimension is based on spatial correlations in phase space, it is
an important precaution to exclude serially correlated points in counting the
pairs (for details see Sect.~\ref{TemporalCorrelations}), and the length of
the window,~$W$, should at least cover all points within the auto-correlation
time (Theiler \cite{Theiler86}). With this correction, the correlation integral
is given by
\begin{equation}
 C(r) =  \frac{2}{\rm N_{pairs}} \sum_{i=1}^{{\rm N}-W} \sum_{j=i+W}^{{\rm N}}
        \Theta(r-\|\mbox{\boldmath{$\xi$}}_i-\mbox{\boldmath{$\xi$}}_j\|) \, , \\
\end{equation}
\begin{equation}
 {\rm N_{pairs}}  =  ({\rm N}-W)({\rm N}-W+1) \, ,
\end{equation}
where N denotes the overall number of data points, and $\Theta$ is the
Heaviside step function. For small distances~$r$, the correlation integral~$C(r)$
is expected to scale with a power of~$r$, and the scaling exponent defines
the correlation dimension~$D_c\,$:
\begin{equation}
 C(r) \propto r^{D_c}\, , \quad {\rm for}\:\, r \to 0 \, ,
 \label{corrdim1}
\end{equation}
\begin{equation}
 D_c=\lim_{r \to 0} \frac{\ln C(r)}{\ln r} \: .
\end{equation}
Practically one computes the correlation integral for increasing embedding
dimension~$m$ and calculates the related $D_c(m)$ in the scaling region.
If the~$D_c(m)$ reach a saturation value~$D_c$ for relatively small~$m$,
this gives an indication that an attractor with dimension~$D_c$ exists underlying
the analyzed time series.

\subsection{Local pointwise dimensions}

The local pointwise dimension is a locally defined variant of the
correlation dimension. Its definition is based on the probability~$p_i(r)$
to find points in a neighborhood of a point~$\mbox{\boldmath $\xi$}_i$
with size~$r\,$:
\begin{equation}
 p_i(r)= \frac{1}{{\rm N}_{\rm pairs}} \sum_{j=1 \atop |j-i| \ge W}^{\rm N}
         \Theta(r-\|\mbox{\boldmath{$\xi$}}_i-\mbox{\boldmath{$\xi$}}_j \|) \:,
 \label{pointdim0}
\end{equation}
where ${\rm N}_{\rm pairs}$ gives the actual number of pairs of points in the sum.
For small distances~$r$, $p_i(r)$ is expected to scale with a power of~$r$, and
the scaling exponent~$D_p(\mbox{\boldmath{$\xi$}}_i)$ gives the local
pointwise dimension at point~$\mbox{\boldmath{$\xi$}}_i$:
\begin{equation}
 p_i(r) \propto r^{D_p(\mbox{\boldmath{$\xi$}}_i)}\, , \quad {\rm for}\:\, r \to 0 \, ,
 \label{pointdim1}
\end{equation}
\begin{equation}%
 D_p(\mbox{\boldmath{$\xi$}}_i)=\lim_{r \to 0} \frac{\ln p_i(r)}{\ln r} \, .
\end{equation}
Averaging~$D_p(\mbox{\boldmath{$\xi$}}_i)$ over all points of the time series
or a number of reference points yields the averaged pointwise dimension,
$\bar{D}_p$, which is equivalent to the correlation dimension and gives a
global description of the geometry of an attractor:
\begin{equation}
\bar{D}_p = \frac{1}{{\rm N}_{\rm ok}} \sum_{i=1}^{{\rm N}_{\rm ok}} D_p(\mbox{\boldmath{$\xi$}}_i) \, ,
\end{equation}
with ${\rm N}_{\rm ok}$ the number of accepted reference points.

\begin{figure*}
 \resizebox{12cm}{11.2cm}{\includegraphics{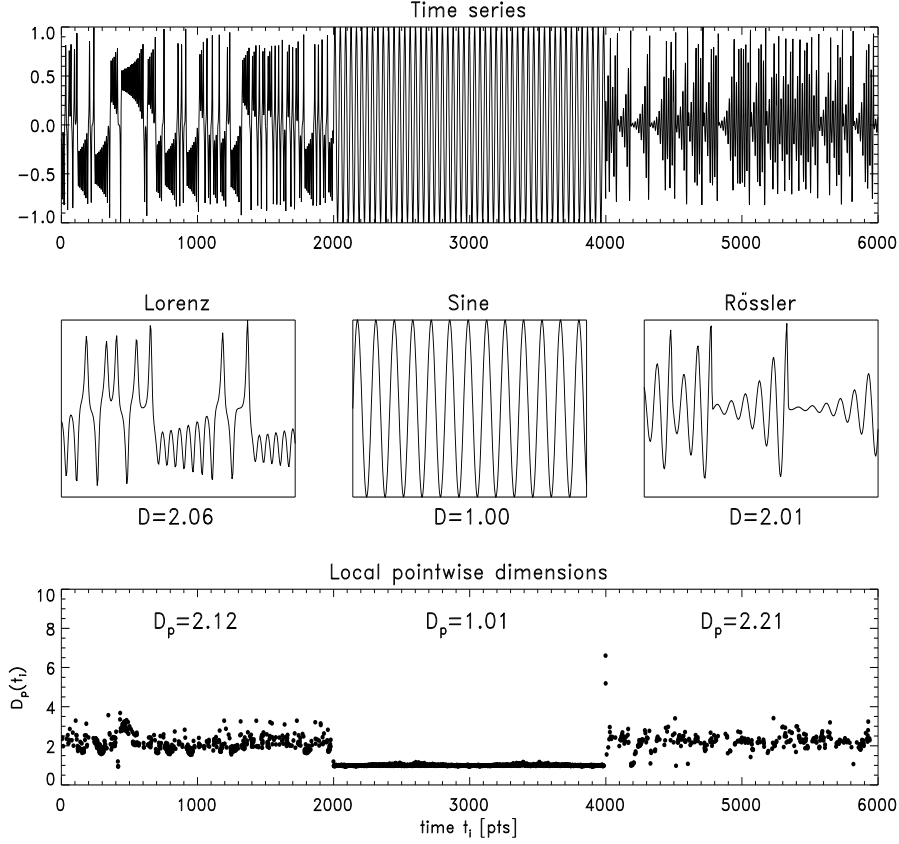}}
 \hfill
 \parbox[b]{55mm}{
 \caption{Top Panel: Simulated time series, made up of three sections
 related to different attractors, the Lorenz, a limit cycle (Sine), and
 the R\"ossler attractor. The middle panels depict an enlargement of the
 respective sections. The bottom panel shows the related local pointwise
 dimensions~$D_p(t_i)$. Its evolution reflects the different attractors
 successively operating in time.}
 \label{9066_f1}
 }
\end{figure*}

Since the $D_p(\mbox{\boldmath{$\xi$}}_i)$ are local functions and defined
for each point, they are, on the one hand, a function of the
position~$\mbox{\boldmath{$\xi$}}_i$ on the attractor, characterizing its
local geometry. On the other hand, the local dimensions can be interpreted as a
function of time, $D_p(t_i)$, since they reflect the temporal evolution, in which
the points~$\mbox{\boldmath{$\xi$}}_i$ on the attractor are covered by the
dynamics (Mayer-Kress \cite{Mayer-Kress94}). Based on this fact, local dimension
estimations have the interesting property that they enable to
cope with non-stationary data.

An example is shown in Fig.~\ref{9066_f1}, which illustrates the local
pointwise dimensions calculated for a simulated time series, made up of
three sections related to different attractors, the Lorenz, a limit cycle
and the R\"ossler attractor. The evolution of the local pointwise
dimensions~$D_p(t_i)$ detects the different attractors successively
operating in time. The averaged pointwise dimensions~$\bar{D}_p$ in the
respective sections deviate less than 10\% from the true values. As for
the calculation of the local dimension at point~$\mbox{\boldmath{$\xi$}}_i$
the distances to {\em all} points of the time series, even those related to
a different attractor, are taken into account, it is a quite striking feature
that the different attractors can be disentangled by the method. Moreover, for
this exemplary analysis the length of the time series was taken rather short
in order to mimic the conditions of observed time series. Such a behavior reveals
that the reference point~$\mbox{\boldmath{$\xi$}}_i$, which is equivalent to a
time~$t_i$ of the dynamical evolution of the system, dominates the local
dimension calculation. However, we want to stress that small changes of the
attractor dimension, as, e.g., in the case of the Lorenz and the R\"ossler
attractor (see Fig.~\ref{9066_f1}), cannot be detected .

\subsection{Pitfalls in dimension estimations}

The determination of fractal dimensions from time series, characterized by
finite length and accuracy, includes many pitfalls, which can lead to quite
spurious results. In this chapter the most prominent problems will be
discussed and the strategies used by the authors to avoid such pitfalls.
A brief but quite dense review with respect to critical points in the
determination of the correlation dimension can be found in Grassberger
et al. (\cite{GrassbergerEtal91}). As the local pointwise dimensions
are a variant of the correlation dimension, most of the problems occur
in a similar way.

\subsubsection{Noise and quality of the scaling region \label{pl_noise}}

As expressed in Eq.~\ref{corrdim1} for the correlation integral~$C(r)$ and
Eq.~\ref{pointdim1} for the probability~$p_i(r)$, respectively, a scaling behavior
is expected for $r \to 0$. However, for real time series, which are contaminated
by noise and which are of finite length, the scaling behavior is expected to
occur at intermediate length scales. Noise is acting at small scales and
therefore dominating the scaling behavior for small~$r$. Deviations from
ideal scaling at large length scales are due to edge effects caused by the
finite length of the time series.

\begin{figure}
 \centering
 \resizebox{0.94\hsize}{!}{\includegraphics{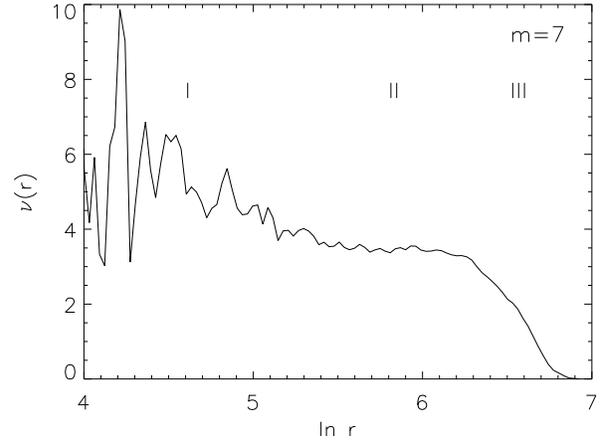}}
 \caption{Scaling behavior for a time series of finite
 length and accuracy, calculated from a stationary subsection of a
 type~IV event with sudden reductions (January 13, 1989).
 On small scales~$r$ [I], the scaling is dominated by noise.
 Since noise tends to fill the whole phase space: $\nu(r,m) \approx m$.
 At intermediate length scales [II], the physically relevant scaling
 is located: $\nu(r,m) \approx D_c(m)$. For large~$r$ [III], deviations
 from the ideal scaling occur due to the finite number of data points. }
 \label{9066_f2}
\end{figure}

Fig.~\ref{9066_f2} shows such typical scaling behavior for a time series
of finite length and accuracy. For better illustration we have plotted
the local slopes of the correlation integral, given by the expression
\begin{equation}
\nu(r) = \frac{{\rm d} \ln C(r)} {{\rm d} \ln r} \, .
\end{equation}
In the case of ideal scaling, the $\nu(r)$, calculated for different
embedding dimensions~$m$, form straight lines parallel to the $x$-axes,
the so-called plateau region, with the constant $y$-value corresponding
to~$D_c(m)$. However, it is typical for observational time series that at
least three different parts in the $\nu(r)$ curves are distinguishable,
described in the caption of Fig.~\ref{9066_f2}. To avoid spurious finite
dimensions which might arise from a misinterpretation of the existence
or non-existence of a physically relevant scaling region, we implemented an
algorithm to automatically check for scaling behavior
(see Sect.~\ref{testalgo}).

\subsubsection{Temporal correlations  \label{TemporalCorrelations} }

The correlation integral~$C(r)$ and the probability~$p_i(r)$ are a measure of
spatial correlations on the attractor. They are basically calculated by counting
pairs of points which are closer to each other than a given distance~$r$. However,
at small scales successive points of the time series give an additional contribution,
since they are close in time. Nevertheless, such points do not reflect spatial
correlations, i.e., a clustering of points in phase space, but are \mbox{serially}
correlated by the temporal order of the time series. To avoid the spurious
contribution of serially correlated points, which can cause strong artificial
effects, especially for data recorded with a high sampling rate, at least all
pairs of points closer than the auto-correlation time have to be excluded
(Theiler \cite{Theiler86}):
\begin{equation}
W > t_{corr} \, .
\end{equation}

A similar phenomenon can occur when the analyzed time series is
rather short. If the time series is too short to ensure that the
attractor is well covered with points, then most of the points of
the series are serially correlated, which again might result in
spurious dimensions. Different relations have been derived for the
minimum length of a time series needed for dimension estimations,
as the one by Eckmann \& Ruelle
(\cite{EckmannRuelle92}):
\begin{equation}
 D_c < 2 \log_{10} {\rm N} \, ,
\end{equation}
with N the length of the time series. However, not only the length of the
data series is of relevance but also the length with regard to the sampling
rate. Applications to different kind of time series, from known chaotic
attractors as well as measured time series (Brandstater \& Swinney
\cite{BrandstaterSwinney87};  Kurths et al. \cite{KurthsEtal91};
Isliker \cite{Isliker92a}; Isliker \& Benz \cite{IslikerBenz94a}), have
revealed that the analyzed time series should at least cover~50
structures -- ``structure" meaning a full orbit in phase space or generally a
typical time scale of the analyzed time series. According to Isliker
(\cite{Isliker92a}), we define the structures by the first minimum of
the auto-correlation function,
\begin{equation}
 {\rm N}_{str}\gtrsim 50  \quad {\rm with} \quad {\rm N}_{str}:= \frac{{\rm N} \cdot \Delta}{t_{corr}} \, ,
 \label{Nstr}
\end{equation}
where ${\rm N}_{str}$ gives the number of structures, N the length of the
time series, $\Delta$ the temporal resolution, and $t_{corr}$ the
auto-correlation time.

\subsubsection{Stationarity}

As shown by Osborne et al. (\cite{OsborneEtal86}) and Osborne \& Provenzale
(\cite{OsborneProvenzale89}), the determination of the correlation dimension
from non-stationary stochastic processes, can erroneously lead to finite
dimension values. To take into account that problem, we applied a stationarity
test proposed by Isliker \& Kurths (\cite{IslikerKurths93}), which is based on
the binned probability distribution of the data. To check for stationarity
one divides the time series into subsections, and compares the probability
distribution of the section under investigation with the probability
distribution of the first half of it by a $\chi^2$-test. By the use of
this test we searched for stationary subsections in the radio burst
time series, and only to such stationary subsections the correlation
dimension analysis was applied.

However, with the concept of local dimensions it is possible to cope
with non-stationary data. One important advantage of the local
dimension method is that it can enable to detect dynamical changes in a time
series. To make use of this potentiality, we calculated the local pointwise
dimensions from the whole time series instead of using stationary subsections.
Moreover, since the statistics in the calculation of the local pointwise
dimensions grows linearly with the length of the time series, whereas
the correlation dimension grows with the square of the number of points,
for the local pointwise dimension analysis the time series should be
kept as long as possible. However, to avoid spurious results due
to non-stationarities, we applied a surrogate data test (Sect.~\ref{surrogate}).

\subsubsection{Intermittency}

Intermittency describes the phenomenon that a time series is
interrupted by quiet phases or phases of very low amplitudes.
Such a phenomenon is quite typical for chaotic systems, but can
cause problematic situations when calculating fractal dimensions
from limited time series. In phase space the intermittent sections
represent regions, which, in the context of the global attractor
scale, degenerate to a point. This would trivially result in an
erroneously low dimension value. One way to cope with this problem
is just to discard intermittent phases from the analyzed time series.
Since the dimension of an attractor is a geometric descriptor, which
means a static quantity, it is not influenced by the serial order of
points, and therefore discarding subsections is a valid strategy.

\begin{figure*}
 \resizebox{12cm}{8.8cm}{\includegraphics{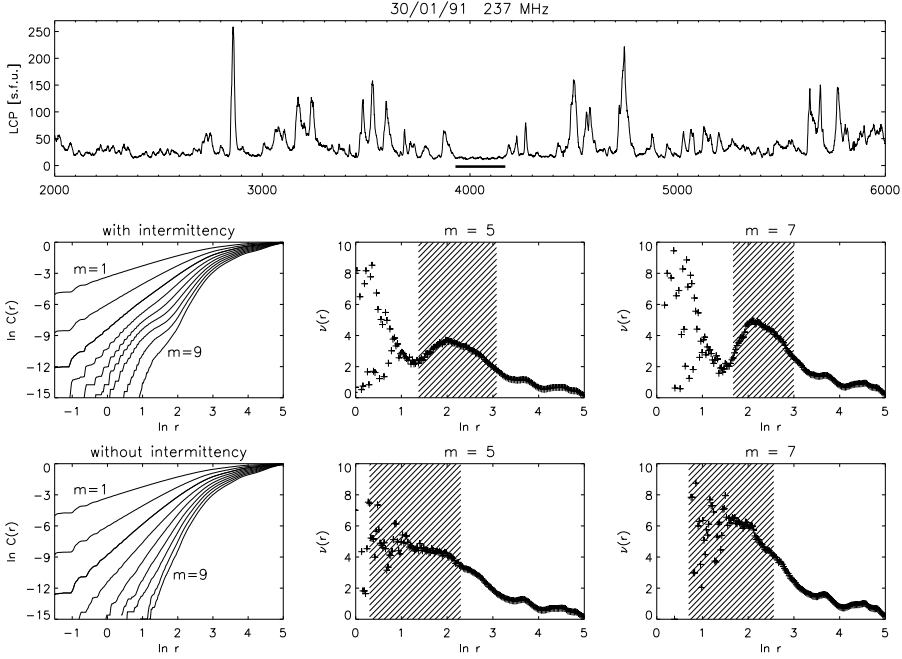}}
 \hfill
 \parbox[b]{55mm}{
 \caption{Top panel: Stationary subsection of a type~I storm
 (January 30, 1991) with an intermittent phase present (marked
 by a line). The abscissa values are given in points (the same
 applies to following figures).
 Middle panel: Correlation integral~$C(r)$ for $m=1$
 to $m=9$, and local slopes $\nu(r)$ for $m=5$ and $m=7$, calculated
 from the whole time series depicted. Bottom panels: Same curves,
 but calculated from the time series after the putative intermittent section
 was eliminated. The hatched ranges in the graphs of the local slopes
 mark the automatically determined scaling region.}
 \label{9066_f3}
  }
\end{figure*}

As an example, the top panel of Fig.~\ref{9066_f3} shows the subsection
of a type~I burst series. The middle and bottom panels show the results of
the correlation dimension analysis, calculated from the whole time series
(middle panels) and the time series after the intermittent section
was eliminated (bottom panels). A comparison of the middle and bottom panels
reveals that the intermittent phase causes a deformation in the curves of the
correlation integral and the related local slopes. Eliminating the
intermittent section, these deformations disappear. However, we want to
stress that the used algorithm for the automatic detection of the scaling
region does not identify the sink-region as a scaling region, which
could result in spurious finite dimensions. Therefore, by the set up
of the scaling algorithm we avoid erroneous low dimensions caused by
intermittency effects.

\subsection{Scaling and convergence test \label{testalgo}}

The practical computation of the local pointwise dimensions turns out to
be more difficult than the correlation dimension, particularly
as for {\em each} reference point the scaling region has to be determined and
the related~$D_p(\mbox{\boldmath{$\xi$}}_i,m)$, which are calculated in the
scaling region, have to be checked for convergence with increasing embedding
dimension~$m$. For this purpose, an automatic and fast procedure is needed.
Moreover, such an automatic procedure can also be used in the correlation
dimension analysis to avoid subjective influences on the scaling and convergence
judgement. We implemented such an algorithm, based on the one used by Skinner
et al. (\cite{SkinnerEtal91}), but modified in order to reach higher stability
and significance. This automatic procedure searches for the scaling region,
defined as the longest linear range in the $\ln p_i(r)$ curves, tests if the
scaling region is of significant length, and finally checks if
the~$D_p(\mbox{\boldmath{$\xi$}}_i,m)$ are converging with increasing~$m$.
Only for points~$\mbox{\boldmath{$\xi$}}_i$, which pass the scaling and the
convergence test, a local pointwise  dimension~$D_p(\mbox{\boldmath{$\xi$}}_i)$
is accepted. In the following we give a description of the algorithm.

\begin{figure}
 \centering
 \resizebox{0.99\hsize}{!}{\includegraphics{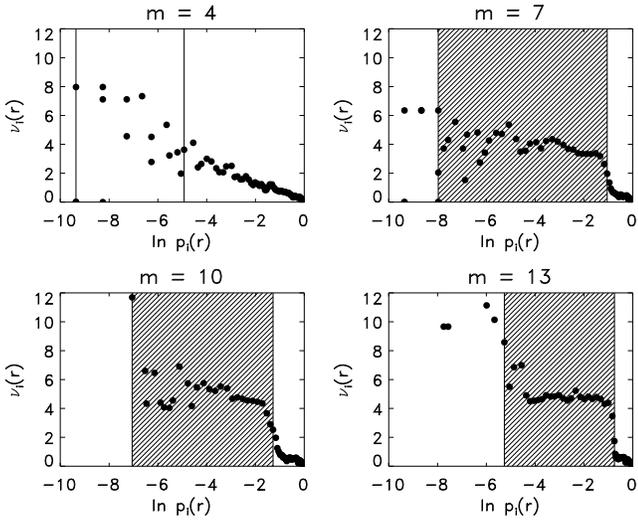}}
 \caption{The curves of local slopes $\nu_i(r)$, with $i=1300$, are shown
 for different embedding dimensions~$m$, calculated from a sample type~I
 storm (August~21, 1991). For  better illustration we plot $\nu_i(r)$
 versus $\ln p_i(r)$ instead of $\nu_i(r)$ versus $\ln r$. The marked regions
 indicate the  location of the scaling range, as detected by the automatic
 procedure. For $m=4$ the determined scaling region is too short (less than
 20\% of the overall $r$-range), and therefore rejected.}
 \label{9066_f4}
\end{figure}

According to Eq.~\ref{pointdim0}, for each reference point~$\mbox{\boldmath{$\xi$}}_i$
and each considered embedding dimension~$m$, we calculate the cumulative histogram
of the~$p_i(r)$, dividing the $r$-range into 100~equidistant points in the
logarithmic representation, which cover the whole range from the smallest to the
largest actual distance. The local slopes of the $\ln p_i(r)$ versus $\ln(r)$ curves,
given by
\begin{equation}
\nu_i(r) = \frac{{\rm d} \ln p_i(r)} {{\rm d} \ln r} \, ,
\end{equation}
calculated in the scaling region, represent
the~$D_p(\mbox{\boldmath{$\xi$}}_i,m)$. For the determination of
the location of the scaling range, we shift a window with length
10~points through the overall $r$-range, and for each of these windows
a least squares linear fit is applied to the $\ln p_i(r)$ versus
$\ln r$ curves. The slope of the linear fit corresponds to the average
value of the local slopes in the corresponding $r$-range, denoted as
$\bar{\nu}_i(r)$. Starting with the first window, successive~$\bar{\nu}_i(r)$
are compared with the corresponding quantity in the first window,
until the difference is larger than a certain threshold value,
chosen as 20\% of the initial~$\bar{\nu}_i(r)$. In this case, the
position of the first point of the start window and the last point
of the actual window are stored, and the procedure continues with the
second window as start window. If the new positions determined cover
a larger range than the old ones, the old values are overwritten by
the new ones, and so on. The two stored values remaining at the end
give the location of the scaling region, in which we
calculate~$D_p(\mbox{\boldmath{$\xi$}}_i,m)$.  If the determined
scaling region is smaller than 20\% of the overall length, it is
interpreted as not significant and rejected. To avoid spurious results
at large~$r$, which usually correspond to small
$\bar{\nu}_i(r)$~values, in the whole procedure we suppress
values $\bar{\nu}_i(r) < 2$. Fig.~\ref{9066_f4} shows a sample
application of the algorithm detecting the scaling range.

After the scaling region is determined for each point~$\mbox{\boldmath{$\xi$}}_i$
and each embedding dimension~$m$, for those points
which reveal a scaling region of significant length for all
considered $m$-values, we test the convergence of
the~$D_p(\mbox{\boldmath{$\xi$}}_i,m)$ with increasing~$m$. This
is simply done by averaging the~$D_p(\mbox{\boldmath{$\xi$}}_i,m)$
over four successive~$m$-values. If the standard deviation turns out
to be less than 15\% of the average value, the convergence is accepted,
and the average of the~$D_p(\mbox{\boldmath{$\xi$}}_i,m)$ over the
considered $m$-range is taken as local pointwise
dimension~$D_p(\mbox{\boldmath{$\xi$}}_i)$. The different kind of
threshold values used have been adjusted by application of the
algorithm to different time series, and slight changes do not
qualitatively change the outcome.

\subsection{Surrogate data test \label{surrogate}}

As already mentioned, local dimensions enable to deal with non-stationary
data. However, to avoid spurious dimensions which can result from non-stationary
stochastic processes, we applied a surrogate data test (Osborne et al.
\cite{OsborneEtal86}; Theiler et al. \cite{TheilerEtal92}). For this purpose,
a Fourier transform of the time series is performed, the Fourier phases are
randomized, and finally to this phase-altered series the inverse Fourier tranform
is applied. The phase randomization keeps the power spectrum unchanged while linear
correlations in the time series are eliminated. If the results of the dimension
analysis of such an ensemble of surrogate data are significantly different from
those computed from the original data, the hypothesis can be rejected that the
obtained results are caused by a linear stochastic process, and nonlinearity
in the data is detected.

To quantify the statistical significance we make use of a hypothesis testing
given in Theiler et al. (\cite{TheilerEtal92}). The null hypothesis assumes that
the original data represent linearly correlated noise. To compute a
discrimination statistics we need a set of surrogate data, from which the same
statistical quantities are derived as from the original time series. In particular,
we use the  averaged pointwise dimensions~as function of the embedding dimension,
$\bar{D}_p(m)$. Let~$Q_o$ denote the
statistical quantity computed for the original time series, and~$Q_{s_i}$ for
the $i$\/th~surrogate generated under the null hypothesis. Let~$\mu_s$ and~$\sigma_s$
represent the (sample) mean and standard deviation of the distribution of~$Q_s$.
With this notation the measure of ``significance",~$S$, is given by the expression
\begin{equation}
S = \frac{|Q_o - \mu_s|}{\sigma_s} \, .
\end{equation}

\begin{table*}[ptbh]
\caption{\label{TableI} Summary of the analyzed type~I storms.
 The first set of columns gives a description of the events, including
 the date and start time in UT, the available number of data points,
 the recording frequency in MHz, and the predominant polarization
 sense (L for Left, R for Right handed circular polarization).
 Each event was recorded with a temporal resolution of 20~ms. In the
 second set of columns, for each event the longest stationary subsection
 (given in number of points) is listed. The third set contains the results
 of the pointwise dimension analysis. We list the mutual information~$t_{mut}$
 in points, the averaged pointwise dimension~$\bar{D}_p$ with standard
 deviation~$\sigma$, the percentage of points passing the scaling and
 convergence test (``ok"), the increase of the averaged pointwise dimension
 with increasing embedding dimension in percent, $\Delta \bar{D}_p$, and
 the outcome of the surrogate data test (pos(itive) means that
 the null hypothesis can be rejected and nonlinearity is detected).}
\begin{tabular}{ccrcc|rr|rccccc} \hline
Date  & Start & \multicolumn{1}{c}{Dur.}  & Frequ. & Pol. & \multicolumn{2}{c|}{Stat. sect. [pts]}             & $t_{mut}$                 & $\bar{D}_p$ & $\sigma$ & ok  & $\Delta \bar{D}_p$ & surr. \rule[0cm]{0cm}{0.35cm} \\
      & [UT]  & \multicolumn{1}{c}{[pts]} & [MHz]  &      & \multicolumn{1}{c}{from} & \multicolumn{1}{c|}{to} & \multicolumn{1}{c}{[pts]} &  [dim]      & [dim]    &[\%] & [\%]               &       \\ \hline
29/04/84 & 11:06:00 & 6000  & 237 & L & 0    & 3600 & 13 & 5.3 & 1.3 & 55 & 7.2  & pos. \\
         & 14:50:30 & 4000  & 237 & L & 200  & 2800 & 14 & 5.6 & 2.0 & 43 & 5.1  & neg. \\
         & 14:54:40 & 10000 & 237 & L & 200  & 7800 & 21 & 6.2 & 1.5 & 50 & 4.1  & pos. \\
25/11/84 & 13:58:55 & 1750  & 237 & R & 550  & 1750 & 5  & 5.1 & 1.3 & 55 & 8.8  & neg. \\
14/05/85 & 11:34:00 & 12000 & 237 & R & 0    & 4000 & 17 & 6.4 & 1.6 & 42 & 6.8  & pos. \\
16/05/85 & 12:51:20 & 8000  & 237 & R & 400  & 5400 & 13 & 6.3 & 1.3 & 52 & 7.7  & pos. \\
06/04/89 & 06:42:00 & 6000  & 237 & L & 0    & 3000 & 9  & 5.1 & 1.3 & 45 & 5.6  & neg. \\
07/04/89 & 11:26:05 & 3000  & 327 & L & 0    & 2800 & 10 & 4.9 & 1.4 & 51 & 5.9  & pos. \\
03/05/89 & 13:33:00 & 5500  & 237 & L & 2400 & 5500 & 12 & 5.3 & 1.6 & 47 & 4.9  & neg. \\
23/06/89 & 10:13:30 & 4500  & 237 & R & 400  & 4000 & 13 & 5.3 & 1.4 & 48 & 6.5  & pos. \\
14/07/89 & 08:48:00 & 3500  & 237 & L & 0    & 3200 & 14 & 5.2 & 1.3 & 55 & 4.8  & pos. \\
19/10/89 & 08:30:20 & 3500  & 237 & L & 0    & 2000 & 6  & 5.4 & 1.5 & 55 & 7.0  & pos. \\
23/11/89 & 12:17:00 & 9000  & 237 & R & 400  & 8400 & 22 & 6.1 & 1.4 & 42 & 5.5  & pos. \\
         & 12:12:20 & 7000  & 327 & R & 400  & 4000 & 14 & 6.2 & 1.3 & 54 & 5.9  & pos. \\
04/01/90 & 11:23:20 & 11000 & 237 & R & 0    & 7800 & 16 & 6.0 & 1.3 & 50 & 5.3  & pos. \\
06/01/90 & 09:04:40 & 3500  & 327 & R & 200  & 3400 & 8  & 5.0 & 1.1 & 52 & 4.6  & pos. \\
01/03/90 & 15:03:40 & 5000  & 237 & L & 1000 & 4000 & 18 & 5.6 & 1.3 & 51 & 5.0  & pos. \\
24/07/90 & 12:42:00 & 12000 & 237 & R & 2000 & 9000 & 21 & 5.7 & 1.1 & 50 & 5.0  & pos. \\
         & 13:13:00 & 5500  & 237 & R & 0    & 2200 & 10 & 5.5 & 1.3 & 52 & 7.0  & pos. \\
22/11/90 & 11:37:10 & 5000  & 237 & L & 0    & 2000 & 12 & 5.9 & 1.4 & 49 & 5.3  & pos. \\
29/01/91 & 12:36:00 & 6000  & 237 & L & 1200 & 6000 & 19 & 5.9 & 1.6 & 46 & 7.7  & neg. \\
30/01/91 & 12:04:20 & 7000  & 237 & L & 600  & 5600 & 23 & 5.9 & 1.4 & 47 & 5.3  & pos. \\
08/05/91 & 10:11:30 & 3000  & 327 & R & 1200 & 2200 & 8  & 5.3 & 1.4 & 48 & 4.7  & pos. \\
10/05/91 & 09:57:30 & 3500  & 327 & R & 60   & 2340 & 9  & 4.9 & 1.1 & 51 & 4.6  & neg. \\
         & 14:16:00 & 8000  & 327 & R & 1000 & 7000 & 14 & 5.8 & 1.6 & 43 & 4.3  & neg. \\
11/05/91 & 10:34:00 & 5000  & 237 & R & 600  & 2500 & 9  & 5.1 & 1.4 & 38 & 4.9  & pos. \\
         & 11:00:20 & 11000 & 237 & R & 1000 &10600 & 18 & 5.6 & 1.4 & 49 & 5.3  & pos. \\
21/08/91 & 07:54:00 & 12000 & 237 & R & 3000 &12000 & 17 & 5.7 & 1.5 & 50 & 4.9  & pos. \\
23/08/91 & 10:47:20 & 7000  & 408 & R & 200  & 7000 & 14 & 5.7 & 1.6 & 50 & 4.9  & pos. \\
28/01/92 & 09:27:20 & 3000  & 408 & R & 200  & 3000 & 10 & 5.7 & 1.3 & 47 & 3.6  & pos. \\ \hline
\end{tabular}
\end{table*}

\begin{table*}[ptbh]
\caption{\label{TableIV} Summary of the analyzed type IV events. The same quantities
 as in Table~\ref{TableI} are listed. Additionally, if particular fine
 structures are present in an event, the predominant type of fine structure is listed
 (pulsations, fast pulsations, sudden reductions, and spikes).}
\begin{tabular}{ccrccl|rr|rccccc} \hline
Date  & Start & \multicolumn{1}{c}{Dur.}  & Frequ. & Pol. & fine str. & \multicolumn{2}{c|}{Stat. sect. [pts]}             & $t_{mut}$                 & $\bar{D}_p$ & $\sigma$ & ok  & $\Delta \bar{D}_p$ & surr. \rule[0cm]{0cm}{0.35cm} \\
      & [UT]  & \multicolumn{1}{c}{[pts]} & [MHz]  &      &           & \multicolumn{1}{c}{from} & \multicolumn{1}{c|}{to} & \multicolumn{1}{c}{[pts]} &  [dim]      & [dim]    &[\%] & [\%]               &       \\ \hline
10/02/84 & 14:44:00 & 4500  & 237 & L &            & 1820 & 4300 & 14 & 6.1 & 1.2 & 40 &   7.1 & pos. \\
14/07/84 & 09:25:00 & 6000  & 237 & R &            & 3000 & 5400 & 12 & 5.8 & 1.4 & 48 &   5.1 & pos. \\
24/04/85 & 10:32:00 & 12000 & 237 & R & fast puls. & 3000 & 10000& 9  & 7.9 & 1.4 & 53 &   7.4 & neg. \\
07/02/86 & 10:50:40 & 10000 & 408 & L & spikes,    & 1000 & 9000 & 8  & 7.0 & 1.3 & 46 &   6.4 & pos. \\
02/01/89 & 10:19:00 & 3000  & 408 & R & sudd. red. & 1050 & 2750 & 7  & 5.1 & 2.1 & 45 &   4.9 & pos. \\
13/01/89 & 12:33:00 & 8000  & 408 & L & sudd. red. & 200  & 2400 & 10 & 5.9 & 2.2 & 54 &   1.1 & pos. \\
12/03/89 & 07:35:50 & 6000  & 327 & L & puls.      & 200  & 6000 & 9  & 6.2 & 1.4 & 45 &   6.3 & pos. \\
30/11/89 & 12:09:20 & 2000  & 610 & L & spikes     & 200  & 1400 & 4  & 6.3 & 2.6 & 27 &   5.4 & pos. \\
19/10/89 & 12:54:30 & 4500  & 610 & R &            & 1000 & 4400 & 10 & 6.4 & 1.9 & 45 &   6.8 & pos. \\
         & 12:58:20 & 4500  & 408 & R &            & 2320 & 3200 & 13 & 5.3 & 1.4 & 50 &   5.2 & pos. \\
19/12/89 & 10:34:10 & 1500  & 610 & L & spikes     & 200  & 1400 & 4  & 5.6 & 2.4 & 40 &   4.7 & pos. \\
27/12/89 & 13:44:05 & 2000  & 408 & R & puls.      & 200  & 2000 & 5  & 7.0 & 1.4 & 46 &   5.6 & neg. \\
         & 13:47:40 & 3500  & 610 & R & sudd. red. & 2450 & 3200 & 3  & 5.6 & 1.8 & 48 &   4.8 & pos. \\
17/04/90 & 14:30:10 & 1250  & 610 & R &            & 100  & 1050 & 4  & 6.0 & 1.3 & 41 &   6.8 & pos. \\
15/05/90 & 13:16:50 & 4000  & 408 & L & spikes     & 0    & 750  & 6  & 6.1 & 1.7 & 56 &   7.2 & pos. \\
02/07/90 & 09:42:00 & 10000 & 237 & L &            & 1000 & 10000& 21 & 7.0 & 1.6 & 38 &   5.7 & neg. \\
27/11/90 & 11:05:30 & 6000  & 327 & R & puls.      & 2000 & 6000 & 9  & 4.6 & 1.3 & 45 &   3.0 & pos. \\
07/03/91 & 08:31:20 & 3500  & 327 & L & puls.      & 0    & 3200 & 12 & 4.8 & 1.3 & 60 &   9.1 & pos. \\
10/07/91 & 12:05:20 & 3000  & 237 & R & puls.      & 0    & 2400 & 12 & 5.7 & 1.8 & 40 &   6.6 & pos. \\
         & 12:04:40 & 10000 & 327 & R & fast puls. & 4000 & 8000 & 7  & 7.0 & 1.5 & 35 &   3.2 & pos. \\
         & 12:04:00 & 12000 & 408 & R & fast puls. & 1000 & 5000 & 8  & 8.2 & 1.3 & 54 &   8.4 & pos. \\
         & 12:04:00 & 12000 & 610 & L & fast puls. & 5000 & 10000& 6  & 7.8 & 1.5 & 48 &   7.9 & neg. \\
         & 12:08:00 & 6000  & 610 & L & fast puls. & 1200 & 4800 & 5  & 7.1 & 1.6 & 47 &   4.7 & pos. \\
22/07/91 & 09:48:00 & 6000  & 610 & R & puls.      & 1200 & 3800 & 10 & 6.1 & 1.4 & 37 &   4.1 & pos. \\
11/11/91 & 12:42:30 & 4500  & 610 & L &            & 200  & 4400 & 8  & 6.0 & 1.5 & 52 &   7.4 & pos. \\
27/02/92 & 11:45:40 & 6000  & 237 & L & fast puls. & 1500 & 3050 & 7  & 7.1 & 1.7 & 49 &   6.4 & pos. \\
         & 11:56:40 & 10000 & 237 & L & fast puls. & 0    & 6000 & 4  & 7.6 & 1.2 & 59 &   8.4 & pos. \\ \hline
\end{tabular}
\end{table*}

\section{Data sets and analysis procedure  \label{DataSets}}

The data sets are single frequency recordings from the multichannel
radio-polarimeter of the Trieste Astronomical Observatory, which is operating in
the dm-m~wavelength range. The investigated data sets are recorded at
the frequencies 237, 327, 408 and 610~MHz, with
a sampling rate of 50~Hz, i.e. a temporal resolution \mbox{$\Delta = 20$~ms}.
We analyzed 30~data sets of type~I storms and 27~sets of type~IV events,
which cover samples with different kind of fine structures, such as pulsations,
fast pulsations, sudden reductions and spikes. Some of the events were
analyzed at different times and/or frequencies. Therefore, the 30~data
sets of type~I storms represent 24~different events, the 27~data sets of
type~IV events 20~different ones. In Table~\ref{TableI} and Table~\ref{TableIV}
we give a description of the type~I and type~IV events, respectively.
The main criteria for the data selection from the solar radio burst
data archive of the Trieste Observatory were:
\begin{enumerate}
\item The selected data sets were representative for the
      particular types of events.
\item To ensure a high signal-to-noise ratio only intense events
      were selected.
\item The related time series were substantially long and
      fulfilled Eq.~\ref{Nstr}.
\end{enumerate}
For the analysis, the predominant polarization sense, LCP (Left-handed
Circular Polarization) or RCP (Right-handed Circular Polarization) of the
burst series was used.

The first step in the correlation dimension analysis was to search
for stationary subsections by shifting windows with decreasing length
through the time series and applying the stationarity test proposed
by Isliker \& Kurths (\cite{IslikerKurths93}). Only those stationary
subsections which still fulfilled the minimum length criterion of Eq.~\ref{Nstr}
were accepted for further analysis, and the correlation dimension was
calculated only from such subsections. The analysis was repeated with
different values for the delay parameter$~\tau$, located around the
first minimum of the mutual information. The relevant quantities were
calculated up to embedding dimension~$m=20$. Finally, the algorithm for
automatically searching the scaling region, checking its validity and
testing the convergence behavior was applied. The convergence was checked
for $m$-intervals containing four successive embedding dimensions,
starting with $m=10$.

The pointwise dimension analysis was basically carried out in the same
manner, except that the overall time series was used instead of stationary
subsections. Moreover, only for points which passed both scaling and
convergence test a local pointwise dimension was accepted. Finally, the
pointwise dimension analysis was repeated for 10 different sets of
surrogate data to test against the null hypothesis that the results are
caused by linearly correlated noise, and to get evidence on nonlinearity
in the data.

\section{Results  \label{Results}}

\subsection{Correlation dimension analysis}

The application of the stationarity test led to the result, that
only two of the preselected events did not reveal stationary subsections
fulfilling Eq.~\ref{Nstr}. These events were excluded from further analysis.
For each of the other events, the longest stationary subsections analyzed
are listed in Table~\ref{TableI} and Table~\ref{TableIV}. If more stationary
subsections were found, for the analysis the longest three were selected,
and non overlapping or only partially overlapping subsections were preferred.
The correlation dimension analysis of the stationary subsections did not
reveal a finite dimension for any of the events. We identify mainly three
cases, in which negative results occurred: {\it no convergence},
{\it no scaling region}, and {\it deformed scaling region}. These three
cases are of course ideal ones, and the practice quite often revealed a
mixture of it. Therefore, for the single event sections analyzed, we do
not specify the different reasons causing the negative result, but describe
and discuss the principal cases in the following subsections in a general frame.

\begin{figure*}
 \resizebox{12cm}{8.0cm}{\includegraphics{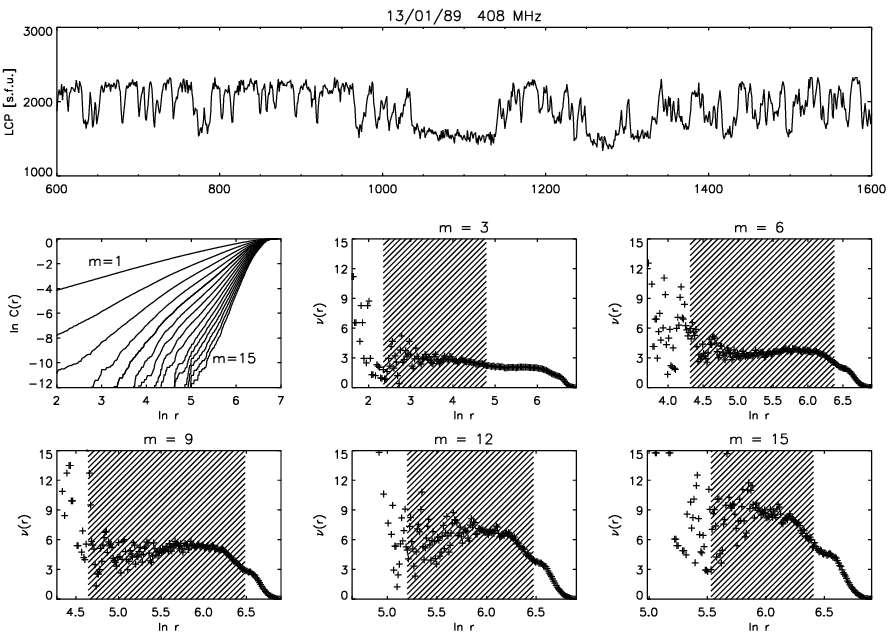}}
 \hfill
 \parbox[b]{55mm}{
 \caption{Top panel: Stationary subsection of a type~IV event
  with sudden reductions (January~13, 1989). The left panel of the
  middle column shows the correlation integral~$C(r)$ for $m=1$
  to $m=15$. The graphs of the local slopes~$\nu(r)$, plotted for
  $m=3$,~6,~9,~12 and~15, reveal distinct plateau regions (hatched),
  which for increasing~$m$ are moving to higher values, indicating
  that the~$D_c(m)$ are divergent.}
 \label{9066_f5}
 }
\end{figure*}

\subsubsection{No convergence}

If the $D_c(m)$ converge with increasing embedding dimension~$m$,
this gives indications for the existence of a low-di\-men\-sional
attractor underlying a time series. In the analyzed data sets,
for which the $C(r)$ curves revealed a clear scaling region, no
convergence with increasing~$m$ occurred. As an example,
Fig.~\ref{9066_f5} shows the divergent behavior for a type~IV event.
Even for rather high embedding dimensions a distinct scaling region
exists. However, as the curves of the local slopes $\nu(r)$ clearly
reveal, the plateau of the scaling region moves to higher values
for increasing~$m$, indicating that the $D_c(m)$ are divergent.
The meaning of such a divergent behavior can be manifold, being
related to the physical state of the system as well as to
restrictions regarding the analysis methodology:
\begin{enumerate}
 \item The underlying physical system is stochastic.
 \item The signal is the output of a deterministic but
  high-dimensional system, with a dimension too high to be
  extracted from the given time series of finite length.
 \item The signal represents a system to which noise is
  coupled intrinsically to the dynamics.
 \item The analyzed time series is the result of different
  physical systems which are independently operating at the
  same time, e.g., different uncoupled radio burst sources
  simultaneously present on the sun, whose emissions sum up
  to the measured signal.
 \item Despite the careful data selection, the measurement noise
  is still too high for the kind of analysis carried out and
  dominates the results.
 \item The choice of the time delay~$\tau$, which is a quite
  critical and sensitive parameter in the correlation dimension
  analysis, is not optimal.
\end{enumerate}

\begin{figure*}
 \resizebox{12cm}{8.0cm}{\includegraphics{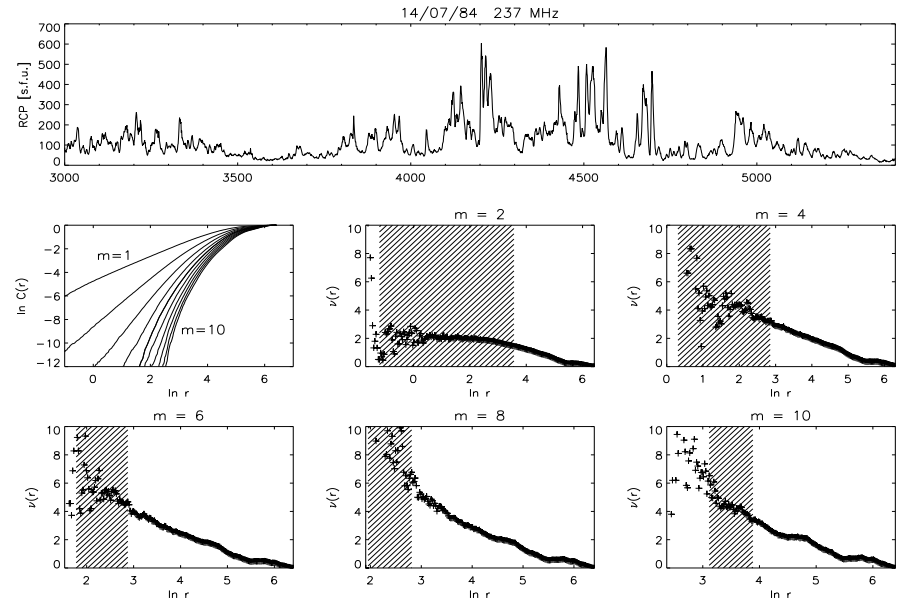}}
 \hfill
 \parbox[b]{55mm}{
 \caption{Top panel: Stationary subsection of a type~IV event
  (July~14, 1984). The curves of the correlation integral~$C(r)$,
  plotted for $m=1$ to $m=10$, reveal only for small~$m$ a linear
  range. In the graphs of the local slopes~$\nu(r)$, the plateau
  region (hatched) shrinks with increasing~$m$, and for $m \gtrsim 6$
  no distinct scaling region exists.}
 \label{9066_f6}
 }
\end{figure*}

\subsubsection{No scaling region}

There exist cases, in which for increasing embedding dimension~$m$
the scaling region disappears. In the curves of the correlation
integral $C(r)$ no linear range can be detected and,
correspondingly, in the graphs of the local slopes~$\nu(r)$ no
plateau region exists. In such cases the quantity~$D_c(m)$ is not
defined. An example is shown in Fig.~\ref{9066_f6}.
The main reason for such a behavior might be that the analyzed
data set is too short and/or too strongly contaminated by noise.
In such a case the deviations from ideal scaling at small and
large length scales~$r$ merge at intermediate~$r$, and cause the
scaling region to disappear. This effect is stronger for higher
embedding dimensions, since for increasing~$m$ the attractor is
sparsely covered with points, which can be verified in Fig.~\ref{9066_f6}.
As shown by Schreiber \& Kantz (\cite{SchreiberKantz95}), even small
amounts of measurement noise can conceal possible scaling behavior.

\subsubsection{Deformed scaling region}

The third case, in which the correlation dimension analysis led to
negative results, is, that the curves of the correlation integral
are highly deformed. Such deviations from ideal scaling are mainly
caused by the presence of intermittent sections, or generally, by
the presence of very different amplitude scales in the data.
Fig.~\ref{9066_f3} shows such deformed scaling regions for a stationary
subsection of a sample type~I storm, caused by an intermittent section.
This can be clarified as the elimination of the intermittent section
causes the deformation to disappear. Explained by the typical presence
of intermittent phases in type~I storms, the correlation dimension
analysis rather often results in deformed plateau regions for type~I
burst series.

\subsection{Local pointwise dimension analysis}

The third set of columns in Table~\ref{TableI} and Table~\ref{TableIV}
contains the results of the pointwise dimension analysis for the type~I
and the type~IV events, respectively. $\bar{D}_p$ gives the averaged pointwise
dimension with standard deviation~$\sigma$, calculated over an embedding
range $m=10-13$, and ``ok" denotes the percentage of points which passed the
scaling and convergence test. $\Delta \bar{D}_p$ gives the increase of the
averaged pointwise dimension, when the embedding range is increased by
two dimensions, from $m=10-13$ to $m=12-15$. ``surr." denotes the outcome of
the surrogate data test, and positive means that the null hypothesis, which
states that the results are caused by linearly correlated noise, can be
rejected, and that significant evidence for nonlinearity in the data is
given.

\begin{figure*}
 \resizebox{12cm}{12cm}{\includegraphics{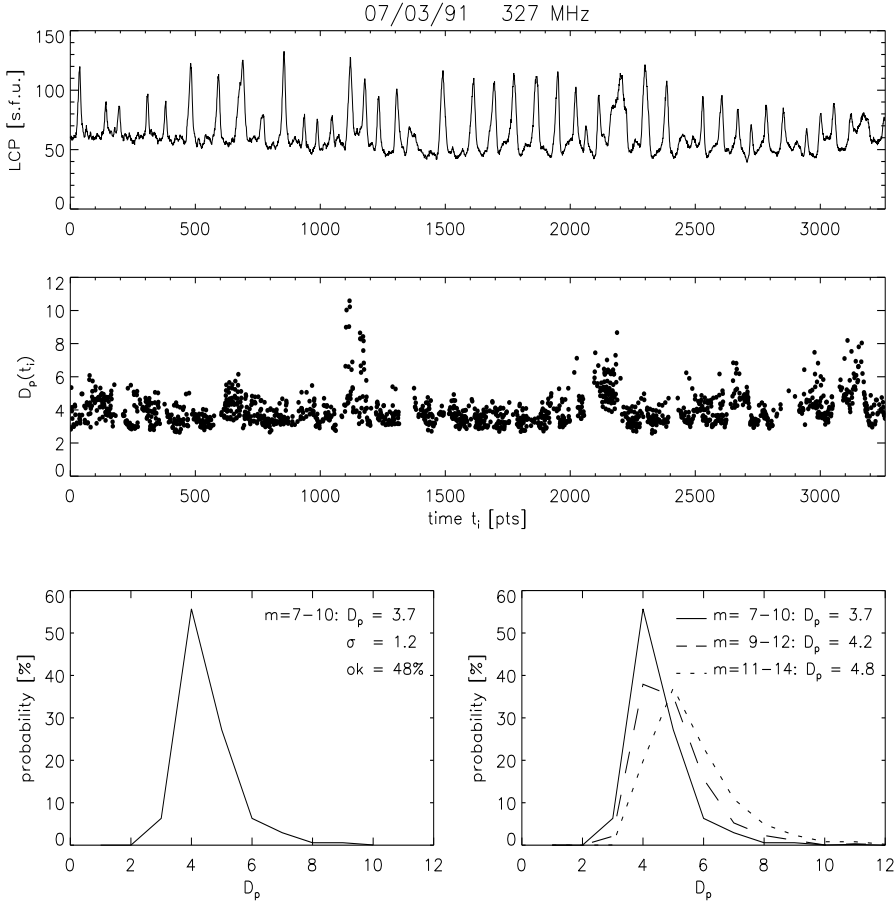}}
 \hfill
 \parbox[b]{55mm}{
 \caption{The top panel depicts the time series of a type~IV event with
  pulsations (March~7, 1991). The middle panel shows the
  evolution of the local pointwise dimensions. The left bottom panel shows
  the histogram of the dimension values. In the right botton panel
  we show the same histogram, overplotted by the histograms of dimensions
  calculated at higher embedding ranges.}
 \label{9066_f7}
 }
\end{figure*}

As an example, Fig.~\ref{9066_f7} illustrates the results of the local
pointwise dimension analysis obtained for a type~IV event with quasi-periodic
pulsations. The top panel shows the time series, the middle panel the time
evolution of the local pointwise dimensions, $D_p(t_i)$. In the bottom
panels the histograms of the local dimensions are plotted, calculated over
different embedding ranges. About half of the points of the time series
passed the scaling and convergence test and were used to compute the averaged
pointwise dimension. However, as the histograms of the local dimensions
calculated over increasing embedding dimensions ($m=7-10$, $m=9-12$, $m=11-14$)
reveal, no absolute convergence exists, but a slight increase with increasing
$m$-ranges occurs. This can be clearly seen in the right bottom panel of
Fig.~\ref{9066_f7}, as for higher embedding ranges the center of the histogram
moves to higher dimension values. The same phenomenon occurs for all analyzed
samples, as the positive $\Delta\bar{D}_p$ values indicate, on the
average $\Delta\bar{D}_p \approx 5\%$.

\begin{figure}
 \centering
 \resizebox{0.999\hsize}{!}{\includegraphics{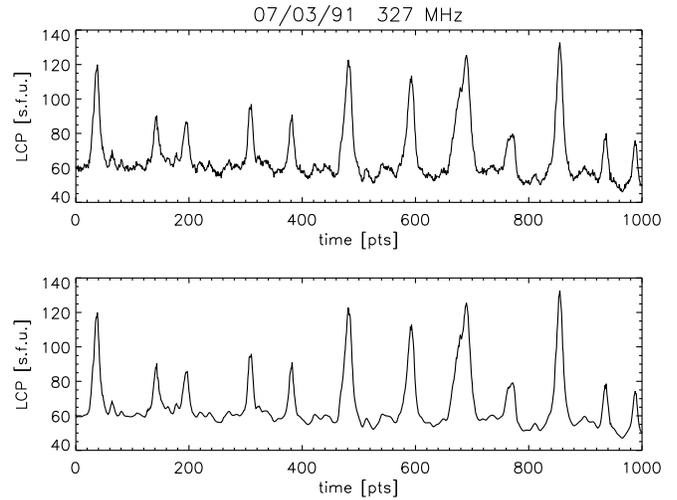}}
 \caption{Illustration of the application of a nonlinear noise reduction.
  The top panel shows the original time series (subsection of a type~IV with
  pulsations, March~7, 1991). The bottom panel shows the same series after
  the application of the noise reduction.}
 \label{9066_f8}
\end{figure}

This phenomenon, on the one hand, could result from the fact that
the~$D_p(\mbox{\boldmath{$\xi$}}_i)$ do not really converge with
increasing~$m$ and no convergence to a low-dimensional attractor
exists. On the other hand, a comparison with simulated time
series from well known chaotic attractors contaminated with Gaussian
noise reveals a similar behavior. For selected samples we repeated the
analysis after application of a simple nonlinear noise reduction
to the data (Schreiber \cite{Schreiber93}), to find out if the increase
is caused by measurement noise. As it is common in nonlinear time \mbox{series}
analysis, the used noise reduction method does not rely on frequency
information but makes use of the structure in the reconstructed phase
space. Fig.~\ref{9066_f8} shows an example of the application of the
noise reduction method. Repeating the analysis with the times series after
the application of the noise reduction led to the effect, that the
dimension increase is softened but not fully eliminated though. Such
result suggests that the dimension increase is not caused by measurement
noise contaminating the signal of a deterministic system.

\begin{figure}
 \centering
 \resizebox{0.95\hsize}{!}{\includegraphics{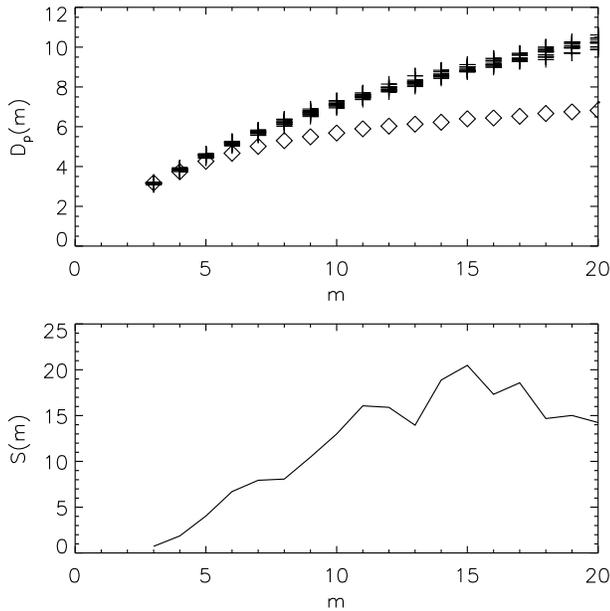}}
 \caption{Illustration of the outcome of the surrogate data test
  for a type~I storm (January~30, 1991). The top panel shows the averaged
  pointwise dimensions, calculated for the original data (diamonds) and for
  an ensemble of 10 surrogates (crosses), for $m=3$ to $m=20$. In the bottom
  panel the significance~$S$ is plotted, which reaches values significantly
  larger than the $3\sigma$~level, giving evidence on nonlinearity in the data.}
 \label{9066_f9}
\end{figure}

The surrogate data analysis yields a positive outcome for most of the
analyzed samples, indicating that the obtained results are not an artifact,
which could arise when the analysis is applied to non-stationary stochastic
data. For $\approx$\,75\% of the type~I and $\approx$\,85\% of the type~IV
burst time series with a significance of~$3\sigma$, we can reject the
null hypothesis that the results are caused by a linear stochastic process,
and have evidence for nonlinearity in the data. Fig.~\ref{9066_f9}
illustrates the outcome of the surrogate data analysis for a sample type~I
storm. The top panel shows the averaged pointwise dimension as a function
of the embedding dimension~$m$, calculated from all points passing the scaling
test. Diamonds represent the original data, crosses the surrogates. The
bottom panel shows the related significance~$S$, which reaches at the
maximum a level of $\approx 20\sigma$, giving strong evidence for nonlinearity in
the data. However, corresponding to the positive $\Delta\bar{D}_p$ values,
the $\bar{D}_p(m)$ of the original data slightly increase with increasing~$m$
and do not show a definitive convergence to a finite dimension value.

\section{Discussion \label{Discussion}}

The outcome of the dimension analysis does not allow to claim low-dimensional
determinism for the analyzed data sets. First, the correlation dimension analysis
failed in all cases. Second, the obtained averaged pointwise dimension
values~$\bar{D}_p$ are too high to characterize a low-dimensional physical
system. Third, the $\bar{D}_p$ do not reveal an absolute convergence with
increasing embedding dimension~$m$. On the other hand though, the local pointwise
dimensions obtained over a specific embedding range yield a quite distinct
behavior, and the surrogate data analysis evidences nonlinearity in the data.

\begin{figure}
 \resizebox{0.99\hsize}{!}{\includegraphics{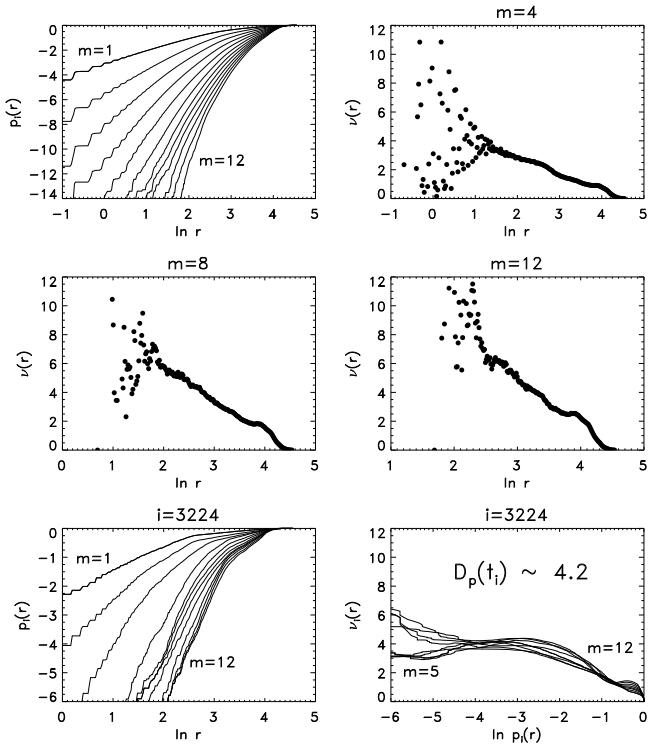}}
 \caption{Comparison of the correlation dimension and the local pointwise
 dimension analysis, calculated for a type~IV burst series with pulsations
 (March 7, 1991). The top and middle panels illustrate the correlation dimension
 analysis. The left top panel shows the curves of the correlation integral~$C(r)$
 for embedding dimensions \mbox{$m=1$} to \mbox{$m=12$}. The top panel at the right
 hand side and the middle panels depict the curves of the local slopes~$\nu(r)$
 for three different embedding dimensions. The bottom panels illustrate the
 outcome of the local pointwise dimension analysis for point $i=3224$. The
 left bottom panel shows the curves of the probability~$p_i(r)$. In the
 right bottom panel we show the curves of the related local slopes $\nu_i(r)$. }
 \label{9066_f10}
\end{figure}

In Fig.~\ref{9066_f10} we show a comparison of the correlation dimension
and the local pointwise dimension analysis of a sample type~IV event. The
correlation integral and the related local slopes do not give any evidence
for low-dimensional determinism, since no significant scaling region and no
convergent behavior occurs. However, the corresponding curves of the local
pointwise dimension analysis reveal a distinct scaling region and a clear convergence
to a finite and low dimension value for certain points~$\mbox{\boldmath $\xi$}_i$.
The figure clearly illustrates that the calculated local dimensions do not
represent an artifact, which might arise, for instance, if the automatic scaling
and convergence procedure is not well adapted to the analysis. Moreover, such
comparison suggests that the local dimension analysis of a times series is more
robust than the classical correlation dimension method. The main reason might be
that in the correlation dimension analysis the scaling behavior itself is a global
property, since {\em all} pairs of points contribute to the correlation integral,
and the scaling region can possibly be smeared out by such an averaging process.
Contrary to that, the local pointwise dimensions are based on the scaling behavior
at {\em each} single point, and can reveal a well defined scaling at certain points.

\begin{table}
\caption{ Statistics of the pointwise dimensions. We list the number of events
 belonging to a particular type or subtype (including type~IV events with fine
 structures of no particular kind, pulsations, fast pulsations, sudden reductions,
 and spikes), the number of events giving a positive outcome of the surrogate data
 test, and the pointwise dimension averaged over the respective (sub)type.
\label{statistic} }
\begin{tabular}{lcccc} \hline
type \rule[-0.2cm]{0cm}{0.6cm}           & number & surr. pos. & $\bar{D}_p^{type}$ & \\ \hline
type I storms      & 30   & 23           & $5.4 \pm 0.4$      & \\
type IV events     & 27   & 23           & $6.4 \pm 1.0$      & \\ \hline
type IV fine str.  & 7    & 6            & $5.9 \pm 0.4$      & \\
type IV puls.      & 6    & 5            & $5.5 \pm 0.7$      & \\
type IV fast puls. & 7    & 5            & $7.4 \pm 0.5$      & \\
type IV sudd. red. & 3    & 3            & $5.6 \pm 0.4$      & \\
type IV spikes     & 4    & 4            & $6.3 \pm 0.6$      & \\ \hline
\end{tabular}
\end{table}

In Table~\ref{statistic} we give a summary of the averaged pointwise
dimensions for the different event types. For the type~IV events we
additionally calculated the quantities separately for the different subtypes.
We list the event type, the number of events belonging to each type (or
subtype), the number of events passing the surrogate test, and the average
of the pointwise dimensions over the respective event types, calculated only
from the samples for which the surrogate data test gave a positive result.
On the average, the type~I storms reveal lower~$\bar{D}_p$ values than the
type~IV events and a significant smaller standard deviation, indicating that
type~I storms represent a comparatively homogeneous class. The large standard
deviation for the type~IV burst series basically reflects the different subtypes.
The average of the particular type~IV subtypes yields $\bar{D}_p$ values
which are significantly different: pulsations and sudden reductions reveal
the lowest values, followed by the type~IVs with fine structures of no
particular kind and spikes, whereas the fast pulsations are characterized
by the highest values.

Such statistics suggests that the~$\bar{D}_p$ are quite well representing the
different types of radio bursts under investigation. This means that even if
low-dimensional determinism cannot be proved, the local dimension analysis
can provide a quantitative description of the events, which is not possible with
the commonly used correlation dimension. As argued by Schreiber (\cite{Schreiber99}),
such description has the drawback that it does not provide an invariant
characterization of a system. On the other hand, it offers an alternative
statistical approach for systems, for which pure determinism cannot be established.
This is of particular interest in astrophysics, since astrophysical systems
represent real-world systems, which cannot be influenced by the observer and
are highly interconnected with their surroundings, making pure determinism rather
improbable.

Moreover, we claim that on a comparative basis the retrieved dimension values
are related to the degree of freedom of the system.
From numerical experiments with known chaotic attractors contaminated with
Gaussian noise we retrieved pointwise dimensions, which are slowly increasing with
increasing embedding dimension~$m$, i.e. the absolute convergence to the definite
attractor dimension disappeared due to the contamination of the deterministic signal
with a stochastic component. This behavior is similar to the one of the analyzed
radio events. However, from this similarity we cannot conclude that the pointwise
dimension analysis is indicative for hidden deterministic chaos in the radio burst
time series, since the retrieved dimension values are too high to characterize
low-dimensional determinism, and for a reliable dimension analysis of high-dimensional
systems much longer time series are needed than we have at disposal (-- a limitation
which is intrinsic to time series representing real-world systems). Nevertheless, what we
can infer from this similarity is that the retrieved dimension values, even if not
representing attractor dimensions, are still indicative for the degree of freedom
of the physical system underlying the time series, characterizing its complexity. Based
on this fact, we make use of the retrieved dimension values to describe the complexity of the
related systems in a comparative way, without claiming or \mbox{supposing the presence of
low-dimensional determinism}.

\section{Conclusions \label{Conclusion}}

In the following items we give a summary of the main results obtained by the presented dimension
analysis of several types of solar radio events, based on the correlation dimension and
the local pointwise dimension method. The results are relevant concerning the physics of the
analyzed events as well as the different methods applied.
\begin{enumerate}
\item The analysis does not enable to claim low-dimensional determinism in the time series.
      This outcome is in agreement with the results obtained by Isliker (\cite{Isliker92b})
      and Isliker \& Benz (\cite{IslikerBenz94a}, \cite{IslikerBenz94b}), who also, among
      others, investigated type~I storms, type~IV events, and spikes. We cannot confirm the
      results of Kurths \& Herzel (\cite{KurthsHerzel86}, \cite{KurthsHerzel87}), Kurths \&
      Karlick$\acute{\rm y}$ (\cite{KurthsKarlicky89}), and Kurths et al. (\cite{KurthsEtal91}),
      who obtained finite dimension values for decimetric pulsations. However, the outcome of
      the present paper does not exclude deterministic chaos in the analyzed time series but
      makes pure low-dimensional determinism, characterized by few free parameters, rather improbable.
\item The analyzed time series are not fully stochastic, i.e. white noise. This fact we infer from
      the distinctly slower increase of the dimension values with increasing embedding dimension
      than expected for fully stochastic processes, which always fill the whole phase space,
      i.e. $D(m) \approx m$. Moreover, the surrogate data analysis suggests that the time series do
      not represent linear stochastic processes.
\item For most of the analyzed data sets we have evidence that nonlinearity in the time
      series is present (given on a $3\sigma$~level by means of a surrogate data test).
\item A comparison of the two different methods used for the determination of fractal dimensions
      reveals that the local dimension method is more stable and enables more physical insight
      than the classical correlation dimension method. The local dimension analysis
      can provide a statistically significant quantity for systems, which cannot be characterized
      by invariants of the dynamics, probably since they are in fact not purely deterministic.
      Such quantities can be of special interest for comparative studies, investigating interrelations
      between different time series (which, e.g., can be useful for classificational purposes) or
      investigating intrarelations in between one time series (in order to detect dynamical changes).
\item The retrieved pointwise dimension values can be interpreted in terms of complexity of the underlying
      physical system. In this frame our analysis indicates that spikes and fast pulsations are the
      signature of systems of higher complexity than pulsations, sudden reductions and type~I storms.
\end{enumerate}

In relation with other kind of analysis of solar radio bursts the presented results might
give further ideas on the physics of the events. In the following we present a short discussion
in this respect, applied to pulsation and spike events, which are quite striking features
associated with solar flares.

Spikes have been intensively studied during recent times. Their short duration and small bandwidth
gives rise to the evidence that they are associated with the energy fragmentation process
in solar flares (Benz \cite{Benz85}, \cite{Benz86}). Based on this connection, Schwarz et al.
(\cite{SchwarzEtAl93}) performed a nonlinear analysis by means of symbolic dynamics methods,
interpreting the spikes appearance in the frequency-time domain as spatio-temporal patterns. This
analysis gives indications that the simultaneous appearance of spikes at different frequencies
is not a purely stochastic phenomenon but may be caused by a non\-linear deterministic (not necessarily
low-dimensional) system or by a Markov process, compatible with a scenario in which spikes at
nearby locations are simultaneously triggered by a common exciter, i.e. the localized sources are
causally connected. In the present paper we find evidence for the spike events analyzed, that they do
not represent a purely stochastic phenomenon in their temporal order either, even if the degree of freedom
of the related physical system is expected to be quite high. Interpreting this result in the frame of the
scenario suggested by Schwarz et al. (1993), it might give indications that the triggering of successive
spikes by a localized source is not caused by a fully stochastic process, but reveals some (possibly
weak) kind of nonlinear causal connection. However, this inference is restricted to the assumption
that the spikes time series rather reflect the physical conditions of the triggering mechanism
than those of the emission.

Pulsations, although a rather marginal phenomenon in the course of solar flares, have reached a wealth
of attention, especially due to the very regular features they sometimes reveal (for a review
see Aschwanden \cite{Aschwanden87}). In previous investigations of the dimensionality of solar pulsations
(Kurths \& Herzel \cite{KurthsHerzel86}, \cite{KurthsHerzel87}; Kurths \& Karlick$\acute{\rm y}$
\cite{KurthsKarlicky89}; Kurths et al. \cite{KurthsEtal91}) the presence of low-dimensional determinism
is reported, with dimensions \mbox{$2.5 \lesssim D \lesssim 3.5$}. Moreover, for one single event a
dynamical evolution from a limit cycle to a low-dimensional chaotic behavior was found (Kurths
\& Karlick$\acute{\rm y}$ \cite{KurthsKarlicky89}). Although we cannot confirm these results,
we want to stress that our analysis suggests that pulsation events, especially quasi-periodic
pulsations, represent the least complex phenomena among the analyzed types of radio events,
i.e. their degree of freedom is expected to be lower than that of other burst types, even if not
low-dimensional. The inferred high-dimensionality and the nonlinear structures detected do not
match with linear MHD oscillation models for pulsations (e.g., Rosenberg \cite{Rosenberg70}; Roberts et al.
\cite{RobertsEtAl84}), in which only a few eigenmodes are excited, but rather favor models of
self-organizing systems of plasma instabilities, which comprise periodic as well as low- and high-dimensional
chaotic behavior. Such a self-organizing model for the electron-cyclotron maser instability, based on
a Lotka-Volterra type equation system, is discussed in Aschwanden \& Benz (\cite{AschwandenBenz88}),
however restricted to limit cycle solutions.

\begin{acknowledgements}
We thank P. Zlobec for helpful discussions and his support in the data
classification. A.V. and A.H. acknowledge the Austrian Ministry of
Sciences. M.M. acknowledges the financial support by ASI and MURST.
\end{acknowledgements}

\end{document}